# Small-Scale Cosmic Microwave Background Anisotropies as a Probe of the Geometry of the Universe

Marc Kamionkowski[1,‡], David N. Spergel[2,*], and Naoshi Sugiyama[3,4,°]

[1] *School of Natural Sciences, Institute for Advanced Study, Princeton, NJ 08540*
[2] *Princeton University Observatory, Princeton, NJ 08544.*
[3] *Department of Astronomy, University of California, Berkeley, CA 94720*
[4] *Department of Physics, Faculty of Science, University of Tokyo, Tokyo 113, Japan*

ABSTRACT

We perform detailed calculations of cosmic microwave background (CMB) anisotropies in a CDM-dominated open universe with primordial adiabatic density perturbations for a variety of reionization histories. We show that to a great extent, the CMB anisotropies depend only on the geometry of the Universe, which in a matter dominated universe is determined by $\Omega$, and the optical depth to the surface of last scattering. In particular, the location of the primary Doppler peak depends primarily on $\Omega$ and is fairly insensitive to the other unknown parameters, such as $\Omega_b$, $h$, $\Lambda$, and the shape of the power spectrum. Therefore, measurements of CMB anisotropies on small scales may be used to determine $\Omega$.

# 1. INTRODUCTION

The recent discovery of anisotropies in the cosmic microwave background (CMB) (Smoot et al. 1992) has ushered in a new era in cosmology (for a recent review, see White, Scott, & Silk 1993). Although the current detections are still accompanied by significant uncertainties, the CMB anisotropies on all angular scales will soon be mapped out with great precision.

The detailed shape of the CMB spectrum can depend quite sensitively on the uncertain parameters involved, and it is possible that detailed measurements of CMB anisotropies can be used to determine some of the uncertain parameters such as the Hubble constant, the baryon density, and spectral index (Bond et al. 1993). In this *Letter*, we point out that CMB anisotropies may also provide information on the value of $\Omega$.

In a previous paper (Kamionkowski & Spergel 1993; hereafter referred to as paper I), we computed the large-angle anisotropies in a CDM-dominated open universe with primordial adiabatic density perturbations. We found that models with $0.35 \lesssim \Omega \lesssim 0.8$ and $0.4 \lesssim h \lesssim 0.6$ (where $h$ is the Hubble constant in units of 100 km sec$^{-1}$ Mpc$^{-1}$) are compatible with the observed amplitude of large-angle CMB fluctuations, the amplitude and spectrum of fluctuations of galaxy counts in the APM, CfA, and QDOT surveys, and an age of the Universe greater than 13 Gyr.

The analytic treatment in Paper I is valid only for angular scales which probe comoving distance scales well outside the horizon at the surface of last scattering. On angular scales smaller than those probed by COBE, peculiar motions of the plasma at the surface of last scattering must be taken into account. An accurate treatment requires numerical solution of the Boltzmann equations.

In the next Section, we discuss the reionization of the Universe by an early generation of star formation and argue that reasonable reionization histories imply an optical depth to Thompson scattering of $\sim 0.5 - 1$ between us and the redshift of recombination. In Section 3, we briefly describe our numerical results, and compare with our previous analytic calculations. We also show that the CMB spectrum depends primarily on $\Omega$ and the optical depth to the surface of last scattering. In the final Section, we make concluding remarks.

# 2. REIONIZATION

In the standard big-bang scenario, most of the plasma in the Universe recombined to form neutral hydrogen and helium at $z \sim 1300$. Yet observations of high redshift quasars at $z \sim 4$ imply that the intergalactic medium has been mostly ionized since at least that redshift (Jenkins & Ostriker 1991). This implies that sometime between $z \sim 4$ and $z \sim 1300$ the Universe was reionized perhaps by a burst of star formation or by early active galactic nuclei. This reionization at redshift $z_R$ has a significant effect on the CMB spectrum (see, e.g. Sugiyama, Silk, & Vittorio 1993) as it suppresses fluctuations at angular scales much smaller than the horizon size at that redshift, $\theta_H \sim 2(\Omega/z_R)^{1/2}$, and produces additional fluctuations both through Doppler scattering at angular scales comparable to $\theta_H$ and at arcsecond to arcminute scales through second order effects (Ostriker & Vishniac 1986, Vishniac 1987).

If we assume that the first stars are similar to those observed today, then we can estimate the number of ionizing photons produced per hydrogen atom turned into stars. Most of the ionizing radiation is produced by massive stars on the main sequence. A 30 $M_\odot$ star burns 7 $M_\odot$ of hydrogen releasing $0.05 M_\odot c^2$ of energy. Roughly 1/4 of this energy is emitted as ionizing radiation (Kurucz 1979). Since these massive stars account for approximately 1/4 of the initial mass function, star formation releases $\sim$ 100 keV in ionizing radiation per processed baryon. For an O star spectrum, for a typical ionization, roughly 13.6 eV goes to ionize hydrogen and an additional 4 eV goes into photoelectric heating. Thus, there are $\sim$ 5000 ionizing photons emitted per processed baryon. This suggests that a minimum of $2 \times 10^{-4}$ of the Universe must be in bound objects for the Universe to be mostly reionized.

These ionizing photons will create a Stromgren sphere of ionized material around each collapsed region. The mass of this Stromgren sphere will be determined by the balance between ionization and $\alpha^{(2)}(T)$, the temperature ($T$) dependent recombination rate to energy levels above the ground state (Spitzer 1978). The number of ionizing photons needed per hydrogen atom to keep the gas mostly ionized at redshift $z$ is therefore, $\sim t(z)/n_e(z)\alpha^{(2)}(T)$, where $t(z)$ is the Hubble time, and $n_e(z)$ is the electron density. For reasonable assumptions about the electron temperature, this implies that $\sim 5 - 20$ photons must be produced per hydrogen atom to keep the Universe reionized back to $\tau \sim 1$. This suggests that the Universe should become reionized when the fraction of mass in bound objects reaches a level of $\sim 2 \times 10^{-4} - 4 \times 10^{-3}$.

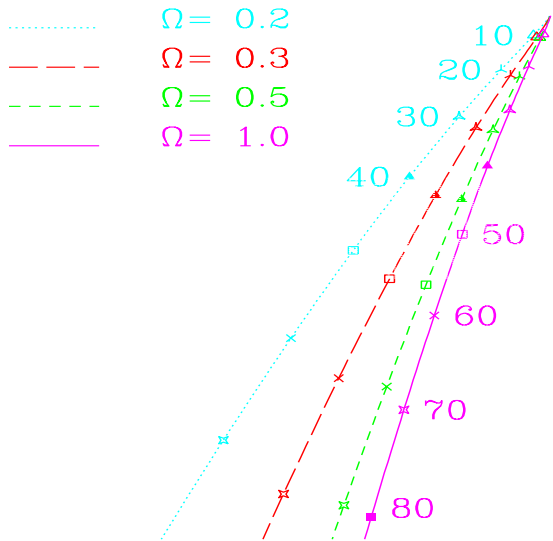

FIG. 1. This figure shows the fraction of the mass of the Universe in collapsed objects as a function of the optical depth between the present and a given redshift. These calculations assumed $h = 0.6$, $\Omega_b h^2 = 0.015$ and a scale-invariant spectrum of density fluctuations. The shaded region corresponds to the fraction of baryons in stars required to ionize the Universe.

We can estimate the fractional mass of the Universe that has collapsed into bound objects of mass greater than the Jeans mass, $f_{coll}(M_J)$, by using the Press-Schechter formalism (Press & Schechter 1974):

$$f_{coll}(M_J, \Omega, z) = \frac{1}{2}\mathrm{Erfc}\left(\frac{1.68}{\sqrt{2}\sigma(M_J,0)D(\Omega,z)}\right). \qquad (2.1)$$

Here, $\sigma(M_J, 0)$ is the amplitude of mass fluctuations today at the Jeans mass scale and $D$ is the linear-theory growth factor. The Jeans mass after recombination is $\sim 10^6 M_\odot$. Fig. 1 shows the fraction of mass in collapsed objects as a function of $\tau(z) \simeq 0.04\Omega_b h\Omega^{-1/2} x_e z^{3/2}$ for different COBE-normalized cosmologies. Using our estimate of required mass fraction for ionization implies that the Universe is reionized at a redshift of 40 – 50 creating an optical depth to Thompson scattering between us and large redshifts of $\sim 0.5 - 1$. Bardeen et al. (1987) argue that unreionized open models violate limits from the OVRO experiment on the amplitude of fluctuations on small angular size. However, the optical depth estimated here is sufficient to smooth out the fluctuations on these scales and evade this limit. As we will see in the next Section, ongoing experiments will enable us to constrain $\tau$ for this family of models.

## 3. RESULTS

The CMB spectra are computed using the numerical techniques described by Sugiyama & Gouda (1992). We assume a scale-invariant Harrison-Peebles-Zel'dovich primordial spectrum of density perturbations. On scales larger than the curvature scale, the definition of scale invariance is somewhat ambiguous, but in Paper I it was shown that this ambiguity affects only the lowest multipole moments ($l \lesssim 9$ for $\Omega \simeq 0.1$ and $l \lesssim 5$ for $\Omega \simeq 0.3$), and that the CMB spectrum is relatively insensitive to the exact shape of the power spectrum on scales larger than the curvature scale.

In Fig. 2 we plot the COBE-normalized CMB spectrum as a function of multipole moment $l$ for several values of $\Omega$ and for optical depths $\tau = 0$ (no reionization) and $\tau = 1$, where $\tau \simeq 0.04\Omega_b h\Omega^{-1/2} x_e(z_{ls})^{3/2}$. Here, $\Omega_b$ is the mass density of baryons, $z_{ls}$ is the redshift of the surface of last scattering, and $x_e$ is the ionization fraction. Note that the location of the first Doppler peak increases as $\Omega$ is decreased, and is insensitive to the ionization history. Reionization decreases the amplitude of the Doppler peak by roughly a factor $e^{-2\tau}$, and leads to another smaller peak at lower $l$ which arises from peculiar motions of scatterers at low redshift in a reionized model. At large $l$, the multipole spectrum drops sharply at the angular scale subtended by the Silk damping scale at recombination. The location of this fall-off is sensitive to $\Omega$.

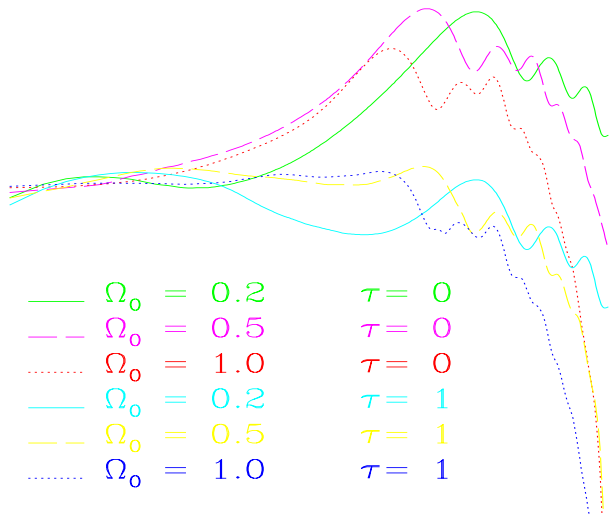

FIG. 2. The COBE-normalized CMB spectrum as a function of multipole moment $l$ for several values of $\Omega$ and for optical depths $\tau = 0$ (no reionization) and $\tau = 1$. Here we have taken $\Omega_b = 0.06$ and $h = 0.5$. Throughout the paper, $\Lambda = 0$.

Interestingly enough, in all these models the spectra on COBE scales, $l \lesssim 13$, are very close to flat, especially when we consider the effect of cosmic variance; at $l = 10$ this introduces a $1\sigma$ error of about 30% in the prediction. The numerical results for the shape and amplitude of the spectrum on these scales agrees well with the analytical results in Paper I, especially for the case of no reionization. On scales $l \gtrsim 15$, the tail end of the Doppler peak becomes important and a numerical calculation is needed. For $\Omega = 1$, the numerical
results are consistent with Bond et al. (1993).

| $\Omega$ | $\tau$ | SP | Sask | MAX | MSAM2 | MSAM3 | ARGO | WD |
|---|---|---|---|---|---|---|---|---|
| 0.2 | 0.0 | 1.1 | 1.1 | 2.1 | 2.3 | 1.6 | 1.3 | 2.0 |
| 0.5 | 0.0 | 1.4 | 1.4 | 2.8 | 3.1 | 2.1 | 1.7 | 1.9 |
| 1.0 | 0.0 | 1.4 | 1.4 | 2.6 | 2.8 | 1.7 | 1.6 | 1.5 |
| 0.2 | 0.5 | 0.8 | 0.8 | 1.3 | 1.5 | 0.9 | 0.8 | 1.1 |
| 0.5 | 0.5 | 1.0 | 1.0 | 1.8 | 1.9 | 1.3 | 1.1 | 1.1 |
| 1.0 | 0.5 | 1.0 | 1.0 | 1.7 | 1.9 | 1.1 | 1.2 | 1.0 |
| 0.2 | 1.0 | 0.8 | 0.8 | 1.0 | 1.2 | 0.6 | 0.8 | 0.7 |
| 0.5 | 1.0 | 1.0 | 1.0 | 1.4 | 1.5 | 0.8 | 1.0 | 0.8 |
| 1.0 | 1.0 | 1.0 | 1.0 | 1.4 | 1.6 | 0.8 | 1.0 | 0.7 |
| 0.2 | 2.0 | 0.9 | 0.8 | 0.9 | 1.1 | 0.4 | 0.8 | 0.4 |
| 0.5 | 2.0 | 1.0 | 1.0 | 1.1 | 1.3 | 0.5 | 0.9 | 0.5 |
| 1.0 | 2.0 | 1.0 | 1.0 | 1.1 | 1.3 | 0.5 | 0.9 | 0.5 |
| Observed | | $0.86 \pm 0.37$ | $1.2 \pm 0.4$ | $4.7 \pm 0.8$ | $3.8 \pm 1$ | $1.8 \pm 0.5$ | $2.1 \pm 0.5$ | < 2.2 |
| | | | | 1.6 | $1.6 \pm 0.5$ | $1.5 \pm 0.4$ | | |

Table 1. Predictions for $(\Delta T/T)_{\rm rms} \times 10^5$ for several experiments for various values of $\Omega$ and $\tau$ and the reported level of observed CMB fluctuations. The two values quoted for the MSAM experiment represent the r.m.s. temperature fluctuations with and without "sources" (Cheng et al. 1993, Page 1993). The two values for the MAX experiment are from two different regions of the sky. The higher value (Gundersen et al. 1993) is reported as a detection. The lower value (Meinhold et al. 1993) may be fluctuation due to free-free emission, synchrotron emission, or may represent CMB fluctuations. For the White Dish experiment (Tucker et al. 1993), we use the "Method I" 2-beam differencing scheme rather than the "Method II" quadrupole differencing scheme. The ARGO (de Bernardis et al. 1993) and SP (Schuster et al. 1993) results are the amplitudes of Gaussian correlation function; the r.m.s. temperature fluctuations will differ slightly from these numbers.
5    6

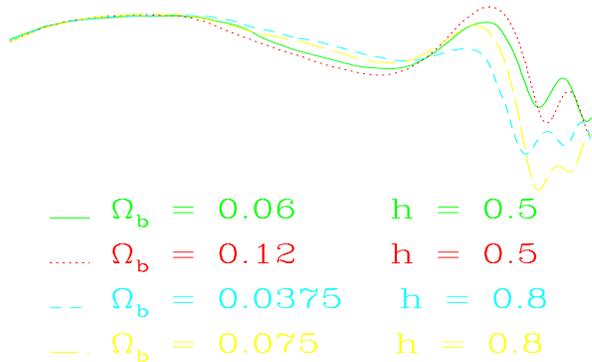

FIG. 3. The COBE-normalized CMB spectrum as a function of multipole moment $l$ for several values of $\Omega_b$ and $h$. In all models, $\Omega = 0.3$ and $\tau = 1$.

In Table 1, we list the predictions for $(\Delta T/T)_{\rm rms}$ for the various models for several different CMB experiments. All of the models are computed for $\Omega_b = 0.06$ and $h = 0.5$. Note that the detection of CMB fluctuations at the level of greater than $10^{-5}$ by the MAX (Meinhold et al. 1993; Gundersen et al. 1993), MSAM (Cheng et al. 1993; Page 1993), SP (Schuster et al. 1993), Saskatoon (Wollack et al. 1993), and ARGO (de Bernardis et al. 1993) experiments suggest that the optical depth of the Universe due to reionization is less than 2. This is consistent with our theoretical expectations for Gaussian theories.

In Fig. 3 we plot the CMB spectrum for several values of $\Omega_b$ and $h$ for $\Omega = 0.3$, and for *fixed* optical depth. The similarity of the curves demonstrates that the primary effect of varying $\Omega_b$, $h$, $z_{ls}$, and $x_e$ can be determined by their scaling with $\tau$. Variation of any of the parameters with fixed $\tau$ has relatively little effect on the CMB spectrum.

Fig. 2 and Fig. 3 show that variations in the location of the Doppler peak are highly insensitive to all the uncertain parameters except for the geometry of the Universe. Although we have used only scale-invariant spectra, the results of Bond et al. (1993) show that the location of the Doppler peak is also relatively insensitive to changes in the primordial spectral index, the addition of a cosmological constant, or the effects of gravitational waves. If the Universe is flat but dominated by a cosmological constant, then the total mass density (non-relativistic matter plus vacuum-energy density) of the Universe is $\Omega = 1$, and the Doppler peak is still located at $l \simeq 200$ (Bond et al. 1993). The robustness of the prediction for the location of the Doppler peak as a function of $\Omega$ should come as no surprise: This simply reflects the angular scale subtended by the horizon at the surface of last scattering.

Information on the value of $\Omega$ should soon be available even without fully mapping out the angular CMB spectrum around the Doppler peak. If $\Omega = 1$, then the amplitude at $l \sim 200$ is greater than that at $l \sim 800$, but if $\Omega < 1$, the ratio is reversed. In Figure 4, we plot the predictions of scale-invariant models with different ionization histories and values of $\Omega$. Figure 4 shows that all of the flat models tend to along a curve in this temperature ratio plot. The various vacuum-dominated and tilted models studied by Bond et al. (1993) also lie along this curve. The position along the curve depends mostly upon the optical depth of the Universe. Open models with $\Omega < 0.5$ lie in upper portion of the plot. This suggests a combination of experiments probing $l \sim 200$ and $l \sim 700$ could potentially determine the geometry of the Universe.

4. CONCLUSIONS

In paper I, we showed that COBE-normalized low-$\Omega$ models are consistent with galaxy counts in large-scale surveys and with the age of the Universe. Due to cosmic variance, as well as the uncertain contribution of tensor modes, discrimination between the various values of $\Omega$ based on measurements of large-angle CMB anisotropies alone was currently impossible and likely to remain so. In this *Letter*, we have performed detailed numerical computations of the CMB spectrum on smaller angular scales and found that information on the value of $\Omega$ *is* most likely encoded in the angular spectrum of CMB anisotropies.

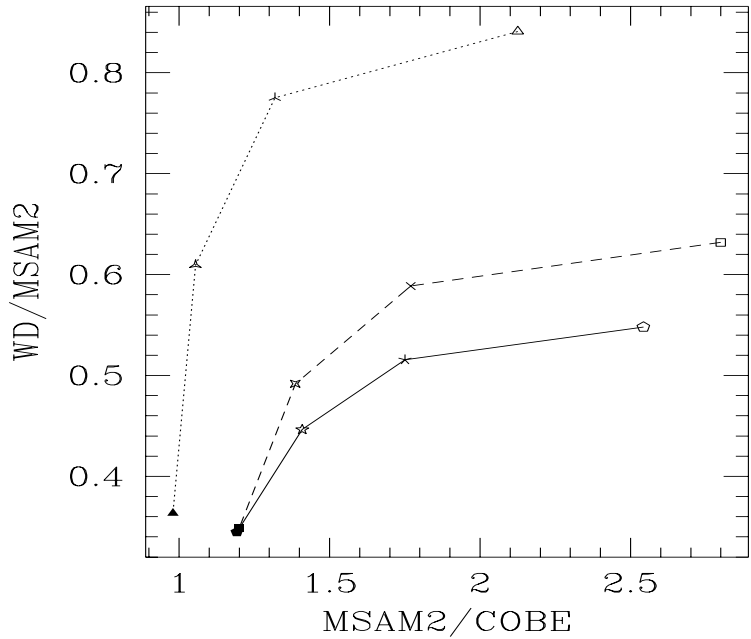

FIG. 4. The points in this plot represent the predicted ratios of temperature fluctuations for different cosmological parameters. The open symbols represent $\tau = 0$ models, the star-shaped symbols represent $\tau = 0.5$ and $\tau = 1$ models, the filled symbols represent the $\tau = 2$ models. The triangular and three-pointed symbols represent $\Omega = 0.2$ models. The square and four-pointed symbols represent $\Omega = 0.5$ models. The pentagons and five-pointed symbols represent $\Omega = 1$ models. The solid, dashed and dotted lines connect the $\Omega = 0.2$, $\Omega = 0.5$ and $\Omega = 1$ models. In the source-free MSAM data set, MSAM2/COBE = $1.6 \pm 0.5$. However, if the sources are included MSAM2/COBE = $3.5 \pm 1.0$. The current White Dish upper limit is not yet strong enough to constrain models.

We have also estimated when the first generation of stars may have reionized the intergalactic medium in these models. Because of the strong dependence of the number of collapse objects on redshift, the predicted epoch of reionization is relatively insensitive to assumption about the details of primordial stars (as long as some massive stars are part of this first stellar generation). In the low $\Omega$ models considered here, we expect that the Universe should have an optical depth of $\sim 0.5 - 1$ between us and the epoch of recombination. This additional opacity to photon scattering reduces the predicted amplitude of fluctuations on small angular scales but does not shift the location of the Doppler peak.

The location of the Doppler peak, $l \sim 200/\Omega^{1/2}$ reflects the angular scale subtended by the horizon at the surface of last scattering, so it should be no surprise that it is relatively insensitive to other cosmological parameters. Similarly, the location of the sharp drop-off in the multipole spectrum at $l \sim 9 \times 10^3 \Omega^{-3/4}(\Omega_b h)^{1/2}$, the angle subtended by the Silk damping scale at the surface of last scatter, also depends on $\Omega$. Although the amplitude of the Doppler peak may be suppressed by reionization, it is quite likely that it will still be distinguishable for reasonable ionization histories.

Current measurements are still too uncertain to discriminate between the various values of $\Omega$. Accurate results from the White Dish and MSAM experiments should be able to distinguish between a flat universe and a universe with $\Omega \sim 0.3$. Experiments with higher resolution which probe multipole moments $100 \lesssim l \lesssim 1000$ (angular scales $0.1° \lesssim \theta \lesssim 1°$) will be able to make the determination more precise. We hope that the ideas presented here will give added impetus to the development of such detectors.

5. ACKNOWLEDGMENTS

We thank T. Herbig, W. Hu, N. Gouda, L. Page, J. Silk, G. Tucker, and M. White for fruitful discussions. We would like to thank L. Page, G. Tucker, and M. White for providing us with appropriate window functions. MK was supported in part by the U.S. Department of Energy under contract DEFG02-90-ER40542. DNS was supported in part by NSF grant AST88-58145 (PYI) and NASA grant NAGW-2448. NS was supported by a JSPS (Japan Society of the Promotion of Sciences) Postdoctoral Fellowship for Research Abroad.